\documentclass[12pt]{JHEP3}
\usepackage{amsmath,amssymb,epsfig}

\font\cmss=cmss12
\newcommand{\vect}[1]{\overrightarrow{#1}}

\newcommand{\bc}{\begin{center}}
\newcommand{\ec}{\end{center}}
\newcommand{\ba}{\begin{array}}
\newcommand{\ea}{\end{array}}
\newcommand{\beq}{\begin{equation}}
\newcommand{\eeq}{\end{equation}}
\newcommand{\bea}{\begin{eqnarray}}
\newcommand{\eea}{\end{eqnarray}}
\newcommand{\bmx}{\begin{pmatrix}}
\newcommand{\emx}{\end{pmatrix}}
\newcommand{\nn}{\nonumber}

\newcommand{\half}{\frac{1}{2}}

\newcommand{\eref}[1]{Eq.~(\ref{#1})}

\newcommand{\cJ}{{\cal J}}
\newcommand{\cN}{{\cal N}}

\newcommand{\tcC}{{\widetilde{\cal C}}}

\newcommand{\vQ}{{\vec{Q}}}
\newcommand{\vP}{{\vec{P}}}

\newcommand{\vA}{{\vec{A}}}
\newcommand{\vB}{{\vec{B}}}
\newcommand{\vC}{{\vec{C}}}
\newcommand{\vD}{{\vec{D}}}

\newcommand{\vr}{{\vec{r}}}

\newcommand{\btau}{{\bar\tau}}
\newcommand{\bX}{{\bar X}}
\newcommand{\bF}{{\bar F}}

\newcommand{\mbps}{m_{\rm BPS}}

\def\IB{\relax{\rm I\kern-.18em B}}
\def\IC{{\relax\hbox{\kern.3em{\cmss I}$\kern-.4em{\rm C}$}}}
\def\ID{\relax{\rm I\kern-.18em D}}
\def\IE{\relax{\rm I\kern-.18em E}}
\def\IF{\relax{\rm I\kern-.18em F}}
\def\II{\relax{\rm I\kern-.18em I}}
\def\IZ{\relax{\sf Z\kern-.35em Z}}
\def\Id{\relax{1\kern-.32em 1}}
\def\IG{\relax\hbox{$\inbar\kern-.3em{\rm G}$}}
\def\IR{\relax{\rm I\kern-.18em R}}
\newcommand\sfrac[2]{{\textstyle\frac{#1}{#2}}}
\newcommand\ncr[2]{\begin{pmatrix}{#1}\\{#2}\end{pmatrix}}
\newcommand\pmat[4]{\begin{pmatrix}{#1}~ &~{#2}\\{#3}~&~{#4}\end{pmatrix}}

\normalsize
\renewcommand{\Im}{{\rm \,Im\,}}
\renewcommand{\Re}{{\rm \,Re\,}}

\title{Constraints on ``rare'' dyon decays} \author{Sunil 
Mukhi\,\footnote{Email: mukhi@tifr.res.in}\,, and Rahul
  Nigam\,\footnote{Email: rahulnig@theory.tifr.res.in}\\ \it Tata
  Institute of Fundamental Research,\\ \it Homi Bhabha Rd, Mumbai 400
  005, India}
\abstract{We obtain the complete set of constraints on 
the moduli of ${\cN=4}$ superstring compactifications that permit
``rare'' marginal decays of $\frac14$-BPS dyons to take place. The
constraints are analysed in some special cases. The analysis extends in
a straightforward way to multi-particle decays. We discuss the
possible relation between general multi-particle decays and
multi-centred black holes.
}

\preprint{TIFR/TH/08-33}


\begin{document}

\section{Introduction}
\label{Introduction}

Recent developments have given us a much better understanding of the
degeneracy counting formula for $\frac14$-BPS dyons in ${\cal N}=4$
string compactifications. This formula has been considerably refined
from its original form in Ref.\cite{Dijkgraaf:1996it} where it was
first proposed. One such refinement consists of specifying the
integration contour in the degeneracy formula and noting that
different contours can lead to different answers for the
degeneracy \cite{Sen:2007vb,Dabholkar:2007vk} (for a review, see
Ref.\cite{Sen:2007qy}). The effect of varying the integration contours is
in the form of discontinuous jumps in the degeneracy whenever the
contour crosses a pole in the integration variable and picks up the
corresponding residue. This has been interpreted as due to the decay
of some $\frac14$-BPS dyons into a pair of $\half$-BPS dyons at curves
of marginal stability, which are computed using the BPS mass formula.

Because for large charges the decaying states are black holes, a
mechanism is needed to explain exactly how these decay on curves of
marginal stability. The answer turns out to be
\cite{Sen:2007pg,Cheng:2007ch} that $\frac14$-BPS black holes (for a
given set of charges) exist both in single-centre and multi-centre
varieties. For the latter, the separations of the centres are
determined by the moduli \cite{Denef:2000nb}. If we specialise to
two-centred dyons with both centres being $\half$-BPS, then it was
shown in Ref.\cite{Sen:2007pg} that as we approach a curve of marginal
stability the two centres fly apart to infinity. On the other side of
the curve the constraint equation has no solutions. This explains (in
principle, though no method is known to explicitly count states of a
two-centred black hole) the phenomenon of marginal stability and
jumping in the counting formula, in terms of the disintegration of
two-centred black holes. It should be noted that the degeneracy of
single-centred black holes with the same charges does not vary across
moduli space, therefore they exist either everywhere or nowhere.  

In these developments, the only type of marginal decay that plays a
role is into two $\half$-BPS final states. Also, the only
multi-centred black holes needed to complete the explanation are those
with a pair of $\half$-BPS centres. Though the correspondence between
these two situations was derived for some special cases, it is
believed to hold in general, namely for any charge vectors and any
point in the entire $\frac{SL(2)}{U(1)} \times
\frac{SO(6,22)}{SO(6)\times SO(22)}$ moduli space of ${\cal N}=4$
compactifications. 

However, there are many more types of marginal decays in the theory,
and in one sense they are far more generic. These decays are into a
pair of $\frac14$-BPS final states, or into three or more final states
each of which can be $\half$-BPS or $\frac14$-BPS. In another sense
these decays are ``rare'', as it has been shown
\cite{Sen:2007nz,Mukherjee:2007nc,Mukherjee:2007af} (at least for unit-torsion
initial dyons) that they take place on curves of marginal stability
that have a co-dimension $>1$ in the moduli space. Therefore these
have been labelled ``rare decays''. In particular they cannot lead to
jumps in the degeneracy formula\footnote{For higher-torsion initial
dyons the curves can be of codimension 1, but the degeneracy (or
rather, index) is still not expected to jump, because of fermion zero
modes. We will focus largely on unit-torsion dyons in this
paper.}. Nevertheless the existence of such decay modes is of
importance in understanding the behaviour of dyons as we move around
in moduli space, and we will study them here for their own sake as
well as for possible interesting physical consequences that they may
turn out to have.

In Ref.\cite{Mukherjee:2007nc} these curves were precisely
characterised as circles on the torus moduli space. But because of
their higher codimension they also need some conditions to be
satisfied in the remaining part of the moduli space. Though the need
for such conditions was demonstrated in
Refs.\cite{Sen:2007nz,Mukherjee:2007nc,Mukherjee:2007af}, the precise
conditions have not yet been worked out. In this paper we will
completely characterise the codimension $>1$ subspace on which rare
decays can take place.

It is also known that there exist multi-centred dyonic black holes
with two $\frac14$-BPS centres, or three or more centres each of which
can be $\half$- or $\frac14$-BPS. However, because the degeneracy
formula does not jump at curves of marginal stability, these
multi-centred dyons have not played a role in studies of dyons in
${\cal N}=4$ compactifications. In particular they have not been
related to marginal decays into two $\frac14$-BPS final states or
multiple final states, and in fact such a relation does not seem
necessary for the state-counting problem. Nevertheless, in what
follows we will argue that the relation between curves of marginal
stability and multi-centred black holes flying apart is quite generic.

In what follows, we start by briefly reviewing what is known about
``rare'' marginal decays in ${\cal N}=4$ compactifications. Then we
find a precise form for the constraints on moduli space in order for
such rare decays to take place. We examine and solve these constraints
in a variety of special cases, to give a flavour of what they look
like. Then using some known results on T-duality orbits, we will
obtain the constraints in the general case. Next we recursively
identify the loci of marginal stability for multi-particle
decays. Finally we examine the special-geometry formula for generic
multi-centred black holes and write it in a form that relates their
separations to curves of marginal stability for $n\ge 2$-body decays.

\section{Marginal stability for ${\cal N}=4$ dyons}

The electric and magnetic charge vectors of a dyon in an ${\cal N}=4$
string compactification are elements of a 28 dimensional integral
charge lattice of signature $(6,22)$. The formulae for BPS mass
involve a $28\times 28$ matrix $L$, which in our basis will be taken
to be:
\beq
\begin{pmatrix}
\label{elldef}
0 ~& \II_6 & 0\\
\II_6 ~& 0 & 0\\
0 ~& 0 & - \II_{16}
\end{pmatrix}
\eeq
as well as a $28\times 28$ matrix $M$ of moduli satisfying 
$MLM^T = L$. The inner product of charge vectors appearing in the BPS
mass is taken with the matrix $L+M$. In the heterotic basis where the
compactification is specified by a constant metric $G_{ij}$, an
antisymmetric tensor field $B_{ij}$ and constant gauge potentials
$A_i^I$ (where $i=1,2,\cdots,6$ and $I=1,2,\cdots,16$), this matrix is
\cite{Maharana:1992my,Sen:1992fr}:
\beq
L + M =
\begin{pmatrix} 
G^{-1} ~&~ 1 + G^{-1} (B + C) ~&~ G^{-1} A \\ 
1 + (-B + C)G^{-1} ~&~ (G-B+C) G^{-1} (G+B+C) ~&~  (G-B+C)G^{-1}A \\ 
A^{T} G^{-1} ~&~ A^{T} G^{-1} (G+B+C) ~&~ A^{T} G^{-1} A
\end{pmatrix}
\eeq
Here $C$ is a symmetric $6 \times 6$ matrix constructed from $A$ as
$C=\half A^T A$, more concretely $C_{ij} =
\frac{1}{2} A_i^I A^I_j$. 

In this basis we parametrise the charge vectors explicitly as:
\beq
\vQ=\begin{pmatrix}
\vQ_{(6)}'\\ \vQ_{(6)}''\\ \vQ_{(16)}'''
\end{pmatrix}
,\qquad
\vP=\begin{pmatrix}
\vP_{(6)}'\\ \vP_{(6)}''\\ \vP_{(16)}'''
\end{pmatrix}
\eeq
where we have broken up the original vectors into three parts with 6,6
and 16 components respectively. In subsequent discussions we will not
explicitly write out the subscripts $(6), (16)$ that appear in the
above formula.

The BPS mass formula for $\frac14$-BPS dyons in ${\cal N}=4$
compactifications is as follows\cite{Cvetic:1995uj,Duff:1995sm}:
\beq
\label{genbpsmass}
\mbps(\vQ,\vP)^2 =
\frac{1}{\sqrt\tau_2}(\vQ-{\bar\tau}\vP)\circ(\vQ-{\tau}\vP) +
2\sqrt{\tau_2}\sqrt{\Delta(\vQ,\vP)} \eeq
where
\beq
\label{Deltadef}
\Delta(\vQ,\vP) \equiv (Q\circ Q)~ (P\circ P) - (P\circ Q)^2
\eeq
The inner products of charge vectors appearing in this formula
are of the form:
\beq
\label{innerprod}
Q\circ P \equiv \vQ^T (L+M)\vP
\eeq
The matrix $L+M$ has 22 zero eigenvalues and therefore the inner
product only contains a projected set of 6 components from the
original 28 components of the charge vector. Explicitly, the zero
eigenvectors take the form:
\beq
\begin{pmatrix}
G + B + C & ~~~A^I~~~ \\ 
-1 & 0 \\ 
 0 & -1
\end{pmatrix}
\eeq
where each column of the above matrix describes an independent zero
eigenvector. 

It is convenient to replace the inner product on charge vectors in
\eref{innerprod} by an ordinary product acting on some projected
vectors. To do this, define $\sqrt{L+M}$ as a $28 \times 28$ matrix
satisfying $\sqrt{L+M}^T \sqrt{L+M}=L+M$. This will be ambiguous upto
a ``gauge'' freedom but we will select a specific solution that is
particularly useful, namely:
\beq
\sqrt{L+M} =
\bmx
~~E^{-1}~~ & ~~E^{-1}(G+B+C)~~ &  ~~E^{-1} A~~\\ 
0 & 0 & 0 \\ 
0 & 0 & 0
\emx
\eeq
With this matrix it is evident that the projected charges only
have their first 6 components nonzero, namely for any arbitrary vectors
$\vQ,\vP$ the projected vectors $\vQ_R,\vP_R$ defined by:
\beq
\vQ_R = \sqrt{L+M}\, \vQ,\qquad 
\vP_R = \sqrt{L+M}\, \vP
\eeq
are 6-component vectors. The components of these vectors are moduli
dependent and not quantised. On the projected vectors, one only needs
to consider ordinary inner products, for example $\vQ_R^T
\vQ_R$ is equal (by construction) to $\vQ^T(L+M)\vQ$. Hence in what
follows we will denote this quantity either by $\vQ\circ \vQ$ or
equivalently by $\vQ_R\cdot \vQ_R$, and analogously for other inner
products.

Within the 6-dimensional projected charge space, the electric and
magnetic charge vectors of the initial dyon span a 2-dimensional
plane. Decay of a $\frac14$-BPS dyon into a set of decay products with
quantised charge vectors
$(\vQ^{(1)},\vP^{(1)}),\cdots,(\vQ^{(n)},\vP^{(n)})$ can take place
only when the plane spanned by the projected charge vectors of
each decay product coincides with this plane (this is the condition
for all states to be mutually $\frac14$-BPS):
\beq
\label{lincomb}
\ncr{\vQ^{(i)}_R}{\vP^{(i)}_R} = 
\pmat{m_i}{r_i}{s_i}{n_i}\ncr{\vQ_R}{\vP_R}
\eeq
When there are just two decay products and both are $\half$-BPS, the
pair of decay products defines a 2-plane. Charge conservation then
implies that this plane coincides with the plane of the original
charge vectors, so in this very special case the above requirement
imposes no conditions on the moduli. Indeed, the numbers
$m_i,r_i,s_i,n_i$ are then integers and the above relation holds
between the full (quantised) charge vectors, not only the projected
ones. Marginal decay takes place on a wall of marginal stability whose
equation is explicitly known (see Ref.\cite{Sen:2007vb} and references
therein). In all other cases, the numbers $m_i,r_i,s_i,n_i$ are
non-integral and moduli-dependent. Then the above condition puts
additional constraints on the background moduli $M$. Our goal here is
to identify these constraints explicitly.

For a two-body decay into $\frac14$-BPS constituents, once the
constraints are satisfied and we find the numbers $m_1,r_1,s_1,n_1$
(the corresponding numbers $m_2,r_2,s_2,n_2$ are determined by
charge conservation) the condition for marginal decay is expressed in
terms of the curve \cite{Mukherjee:2007nc}:
\beq
\label{cms}
\left(\tau_1-\frac{m_1- n_1}{2s_1}\right)^2
+ \left(\tau_2 + \frac{E}{2s_1}\right)^2
= \frac{1}{4s_1^2}\Big((m_1-n_1)^2 + 4r_1s_1 + E^2\Big)
\eeq
Here we have restricted to the case of unit-torsion dyons, so we have
put $m=n=1$ with respect to the notation in
Ref.\cite{Mukherjee:2007nc}.  Also, $E$ is defined by:
\beq
E\equiv \frac{1}{\sqrt\Delta}\left(\vQ^{(1)}\circ \vP - \vP^{(1)}\circ
\vQ\right)
\eeq
Interestingly the numerator of this quantity is the Saha angular
momentum between one of the final-state dyons and the initial state,
evaluated with respect to the moduli at infinity. Exchanging the role
of the two final-state dyons sends $E\to -E$. It also sends
$m_1-n_1\to (1-m_1)-(1-n_1)=-(m_1-n_1)$ and $r_1,s_1\to
-r_1,-s_1$. The curve of marginal stability is invariant under this
set of transformations, as it should be.

We now turn to the detailed study general two-body decays into
$\frac14$-BPS constituents. We will find explicit expressions for the
numbers $m_1,r_1,s_1,n_1$ in terms of the quantised charge vectors
$\vQ,\vP,\vQ_1,\vP_1$ and the moduli $M$. We will also explicitly
characterise the loci in moduli space where such rare decays are
allowed.

\section{Rare dyon decays}

\subsection{Analysis and implicit solution}

It will be useful to define a quartic scalar invariant of four
different vectors by:
\beq
\Delta(\vA,\vB;\vC,\vD) \equiv \det 
\begin{pmatrix}
~\vA\circ \vC ~&~\vA\circ \vD~\\
~\vB\circ \vC ~&~\vB\circ \vD~
\end{pmatrix} =
(\vA\circ \vC)(\vB\circ \vD) - 
(\vA\circ \vD)(\vB\circ \vC)
\eeq
As explained above, the ``$\circ$'' product is the moduli-dependent
inner product involving the matrix $L+M$. The above quantity is
antisymmetric under exchange of the first pair or last pair of
vectors, and symmetric under exchange of the two pairs.  The quartic
invariant of two variables defined earlier is a special case of this
new invariant:
\beq
\Delta(\vQ,\vP) = \Delta(\vQ,\vP;\vQ,\vP)
\eeq

Now start with the following vector equation that is part of
\eref{lincomb}:
\beq
\vQ^{(1)}_R = m_1 \vQ_R + r_1 \vP_R
\eeq
Contracting this successively with $\vQ_R$ and $\vP_R$ we find:
\bea
\vQ^{(1)}_R\cdot \vQ_R &=& m_1 \vQ_R^2 + r_1 \vQ_R\cdot \vP_R\nn\\
\vQ^{(1)}_R\cdot \vP_R &=& m_1 \vQ_R\cdot \vP_R  + r_1 \vP_R^2
\eea
Multiplying the first equation by $\vP_R^2$ and the second by
$\vQ_R\cdot \vP_R$ and subtracting, we find:
\beq
m_1 \Delta(\vQ_R,\vP_R) = \Delta(\vQ_R,\vP_R;\vQ^{(1)}_R,\vP_R)
\eeq 
which enables us to solve for $m_1$. Repeating this process we can
solve for  $r_1,s_1,n_1$ leading to the result:
\beq
\pmat{m_1}{r_1}{s_1}{n_1} =
\frac{1}{\Delta(\vQ_R,\vP_R)}
\pmat{~\Delta(\vQ_R,\vP_R;\vQ^{(1)}_R,\vP_R)~}%
{~\Delta(\vQ_R,\vP_R;\vQ_{R},\vQ^{(1)}_{R})~}%
{~\Delta(\vQ_R,\vP_R;\vP^{(1)}_{R},\vP_R)~}%
{~\Delta(\vQ_R,\vP_R;\vQ_R,\vP^{(1)}_{R})~}
\eeq

It follows that \eref{lincomb} can be expressed as:
\beq
\label{lincombtwo}
\ncr{\vQ^{(1)}_{R}}{\vP^{(1)}_{R}} = 
\frac{1}{\Delta(\vQ_R,\vP_R)}
\pmat{~\Delta(\vQ_R,\vP_R;\vQ^{(1)}_{R},\vP_R)~}%
{~\Delta(\vQ_R,\vP_R;\vQ_{R},\vQ^{(1)}_{R})~}%
{~\Delta(\vQ_R,\vP_R;\vP^{(1)}_{R},\vP_R)~}%
{~\Delta(\vQ_R,\vP_R;\vQ_R,\vP^{(1)}_{R})~}
\ncr{\vQ_R}{\vP_R}
\eeq
For fixed, quantised charge vectors $\vQ,\vP$ of the initial dyon and
$\vQ^{(1)},\vP^{(1)}$ of the first decay product (the charge of the
second product is determined by charge conservation), the above
equation provides a set of constraints on the moduli that must be
satisfied for the $\frac14\to \frac14 + \frac14$ decay to be
possible. These constraints together with the curve of marginal
stability \eref{cms} provide a necessary and sufficient set of
kinematic conditions for marginal decay.

In the above form, it is rather difficult to disentangle the
constraints or to physically understand their significance. Therefore
we will consider a number of special cases. Along the way we will see
the advantage of using T-duality to bring the charges into a
convenient form and performing the analysis in that basis. Finally we
write down the explicit constraint equation in the general case, again
in the chosen T-duality basis.

\subsection{Explicit solution: special cases}

\subsubsection*{(i) $\half$-BPS final states}

The case where the decay products are $\half$-BPS should provide no
constraints on the moduli as this is a ``non-rare'' decay. This
provides a check on our equations. Inserting the $\half$-BPS
conditions:
\beq
\vP^{(1)}=k_1\,\vQ^{(1)},\quad \vP^{(2)}=k_2\,\vQ^{(2)}
\eeq
we find that:
\beq
\pmat{m_1}{r_1}{s_1}{n_1} =
(k_2-k_1)\frac{\Delta(\vQ^{(1)}_{R},\vQ^{(2)}_{R})}{\Delta(\vQ_R,\vP_R)}
\pmat{k_2}{-1}{k_1k_2}{-k_1}
\eeq
We also have:
\beq
\Delta(\vQ_R,\vP_R)= (k_2-k_1)^2 \Delta(\vQ^{(1)}_{R},\vQ^{(2)}_{R})
\eeq
Substituting in the above equation, we find:
\beq
\pmat{m_1}{r_1}{s_1}{n_1} =
\frac{1}{k_2-k_1}\pmat{k_2}{-1}{k_1k_2}{-k_1}
\eeq
At this stage all moduli-dependence has disappeared from the matrix,
and equation \eref{lincomb} indeed provides no constraints on the
moduli. Rather, it reduces to an identity. It is also easy to see that
$k_1-k_2$ divides the torsion of the original dyon, so if we are also
considering the unit-torsion case then $k_1-k_2=1$ and
$m_i,r_1,s_1,n_1$ are all manifestly integral \cite{Sen:2007vb}.

\subsubsection*{(ii) Special charges and moduli}

The next special case we will study has a restricted set of charges.
Additionally, some of the background moduli are set to a specific
value, namely zero in the chosen coordinates. We then examine the
constraints on the remaining moduli. In choosing special values for
the moduli, we should in principle avoid loci of enhanced gauge
symmetry where the dyons we are studying would become massless.

Let us restrict ourselves to special initial-state charges
given by:
\beq
\vQ' = (Q_1',0,\cdots,0),\qquad
\vQ'' = (Q_1'',0,\cdots,0),\qquad
\vQ ''' = 0
\eeq
and 
\beq
\vP' = (0,P_2',0,\cdots,0),\qquad
\vP'' = (0,P_2'',0,\cdots,0),\qquad
\vP ''' = 0
\eeq

Next we set $B_{ij}=0=A^I_i$ as well as $G_{ij}=0,~i\ne j$. 
The above restrictions allow us to choose the orthonormal frames
$E_{ai}$ to be diagonal:
\beq
\label{diage}
E_{ii}= R_i,~i=1,2,\cdots,6
\eeq
with $R_i$ the radii of the six compactified directions in the
heterotic basis. 

In the restricted subspace of moduli space that we are considering
here, the matrix $\sqrt{L+M}$ reduces to:
\beq
\sqrt{L+M} =
\bmx
~~E^{-1}~~ & ~~E~~ & ~~0~~ \\ 
0 & 0 & 0 \\ 
0 & 0 & 0
\emx
\eeq
with $E$ given as in \eref{diage}. Therefore the projected
initial-state charge vectors are:
\beq
\vQ_R = 
\bmx
\frac{Q_1'}{R_1} + Q_1'' R_1\\
0\\ ...\\ 0
\emx, \qquad
\vP_R = 
\bmx
0\\ \frac{P_2'}{R_2} + P_2'' R_2\\
0\\ ...\\ 0
\emx
\eeq
For this configuration we clearly have $\vQ_R\cdot \vP_R=0$ and
therefore the quartic invariant $\Delta$ is:
\beq
\Delta(Q_R,P_R) = \left(\frac{Q_1'}{R_1} + Q_1'' R_1\right)^2
\left(\frac{P_2'}{R_2} + P_2'' R_2\right)^2
\eeq

We take the decay products to have generic charges
$\vQ^{(1)},\vP^{(1)}$ and $\vQ^{(2)},\vP^{(2)}$ subject of course to the
requirement that they add up to $\vQ,\vP$. We then have:
\beq
\vQ^{(1)}_R = 
\bmx
\frac{Q^{(1)'}_1}{R_1} + Q^{(1)''}_1 R_1\\
\frac{Q^{(1)'}_2}{R_2} + Q^{(1)''}_2 R_2\\
...\\
\frac{Q^{(1)'}_6}{R_6} + Q^{(1)''}_6 R_6\\
\emx,~~
\vP^{(1)}_R = 
\bmx
\frac{P^{(1)'}_1}{R_1} + P^{(1)''}_1 R_1\\
\frac{P^{(1)'}_2}{R_2} + P^{(1)''}_2 R_2\\
...\\
\frac{P^{(1)'}_6}{R_6} + P^{(1)''}_6 R_6\\
\emx
\eeq
Now we can compute the quartic invariants appearing in
\eref{lincombtwo}: 
\bea
\Delta(\vQ_R,\vP_R;\vQ^{(1)}_{R},\vP_R) &=&
\left(\frac{Q_1'}{R_1} + Q_1'' R_1\right)
\left(\frac{Q^{(1)'}_1}{R_1} + Q^{(1)''}_1 R_1\right)
\left(\frac{P_2'}{R_2} + P_2'' R_2\right)^2\nn
\\
\Delta(\vQ_R,\vP_R;\vQ_{R},\vQ^{(1)}_{R}) &=&
\left(\frac{Q_1'}{R_1} + Q_1'' R_1\right)^2
\left(\frac{Q^{(1)'}_2}{R_2} + Q^{(1)''}_2 R_2\right)
\left(\frac{P_2'}{R_2} + P_2'' R_2\right)\nn
\\
\Delta(\vQ_R,\vP_R;\vP^{(1)}_{R},\vP_R) &=&
\left(\frac{Q_1'}{R_1} + Q_1'' R_1\right)
\left(\frac{P^{(1)'}_1}{R_1} + P^{(1)''}_1 R_1\right)
\left(\frac{P_2'}{R_2} + P_2'' R_2\right)^2\nn
\\
\Delta(\vQ_R,\vP_R;\vQ_R,\vP^{(1)}_{R})&=&
\left(\frac{Q_1'}{R_1} + Q_1'' R_1\right)^2
\left(\frac{P^{(1)'}_2}{R_2} + P^{(1)''}_2 R_2\right)
\left(\frac{P_2'}{R_2} + P_2'' R_2\right)\nn
\eea
Had we not taken $E$ to be diagonal, the expressions above would have
quickly become very complicated to write down.

Inserting the above expressions, and cancelling some common factors, 
the constraint equation \eref{lincombtwo} becomes:
\bea
\left(\frac{Q_1'}{R_1} + Q_1'' R_1\right)
\left(\frac{P_2'}{R_2} + P_2'' R_2\right)\vQ^{(1)}_R &=&
\left(\frac{Q^{(1)'}_1}{R_1} + Q^{(1)''}_1 R_1\right)
\left(\frac{P_2'}{R_2} + P_2'' R_2\right)\vQ_R +\nn\\
&&\left(\frac{Q_1'}{R_1} + Q_1'' R_1\right)
\left(\frac{Q^{(1)'}_2}{R_2} + Q^{(1)''}_2 R_2\right)\vP_R\nn\\
\left(\frac{Q_1'}{R_1} + Q_1'' R_1\right)
\left(\frac{P_2'}{R_2} + P_2'' R_2\right) \vP^{(1)}_R &=&
\left(\frac{P^{(1)'}_1}{R_1} + P^{(1)''}_1 R_1\right)
\left(\frac{P_2'}{R_2} + P_2'' R_2\right)\vQ_R +\nn\\
&&\left(\frac{Q_1'}{R_1} + Q_1'' R_1\right)
\left(\frac{P^{(1)'}_2}{R_2} + P^{(1)''}_2 R_2\right)\vP_R\nn\\
\eea
These are 6+6 equations. However, the first two components of each set
are identically satisfied, as one can easily check. This is expected,
and follows from the structure of \eref{lincomb} from which
$m_1,r_1,s_1,n_1$ were determined. The remaining four components of
each equation give the desired constraints on the
moduli. Because of the way we have chosen $\vQ,\vP$, the RHS already
vanishes on components 3 to 6, so the constraint is simply that the
LHS vanishes. That in turn sets to zero the components 3 to 6 of the
vectors $\vQ^{(1)}_R$ and $\vP^{(1)}_R$. Thus we find the constraints:
\bea
\label{constr}
\frac{Q^{(1)'}_i}{R_i} + Q^{(1)''}_i R_i =0,\quad i=3,4,5,6\nn\\
\frac{P^{(1)'}_i}{R_i} + P^{(1)''}_i R_i =0,\quad i=3,4,5,6
\eea
If the components of $\vQ^{(1)},\vP^{(1)}$ are all
nonvanishing, this implies that:
\beq
R_i = \sqrt{-\frac{Q^{(1)'}_i}{Q^{(1)''}_i}}
= \sqrt{-\frac{P^{(1)'}_i}{P^{(1)''}_i}},\quad i=3,4,5,6
\eeq
In this special case the constraint equations have some particular
features. First of all, for generic charge vectors $\vQ^{(1)}$ and
$\vP^{(1)}$, there are no solutions. To have any solutions at all, one
must choose the charges of the decay products in such a way that the
second equality in the above equation can be satisfied. In other
words, the sign of $Q_i^{(1)'}$ and $Q_i^{(1)''}$ must be opposite
(for $i=3,4,5,6$), and the same has to be true for $P^{(1)}$. In this
case we find four constraints on the moduli, which fix the
compactification radii $R_3,R_4,R_5,R_6$.

For this special case, the numbers $m_1,r_1,s_1,n_1$ in \eref{lincomb}
are given by:
\bea
m_1 &= \frac{\frac{Q^{(1)'}_1}{R_1} 
+ Q^{(1)''}_1 R_1}{\frac{Q_1'}{R_1} +
Q_1'' R_1}, \qquad
r_1 &= \frac{\frac{Q^{(1)'}_2}{R_2} 
+ Q^{(1)''}_2 R_2}{\frac{P_2'}{R_2} +
P_2'' R_2} \nn \\
s_1 &= \frac{\frac{P^{(1)'}_1}{R_1} 
+ P^{(1)''}_1 R_1}{\frac{Q_1'}{R_1} +
Q_1'' R_1}, \qquad
n_1 &= \frac{\frac{P^{(1)'}_2}{R_2} + P^{(1)''}_2 R_2}{\frac{P_2'}{R_2} +
P_2'' R_2} 
\eea
We see that $m_1,s_1$ depend only on $R_1$ and $r_1,n_1$ depend only
on $R_2$.

So far the decay products were taken to have generic charges
(consistent of course with charge conservation). The situation changes
if we choose less generic decay products.  Earlier we took
all components of $\vQ^{(1)},\vP^{(1)}$ are nonvanishing. However
if $Q^{(1)'}_i=Q^{(1)''}_i=P^{(1)'}_i=P^{(1)''}_i=0$ for any $i\in
3,4,5,6$ then the corresponding constraint \eref{constr} is trivially
satisfied. In this situation we will have a reduced number of
constraints. As an example if the above situation holds for all
directions except $i=3$ and if $\frac{Q^{(1)'}_3}{Q^{(1)''}_3}
=\frac{P^{(1)'}_3}{P^{(1)''}_3}$ then there is only a single
constraint coming from the above equations. The curve of marginal
stability provides one more constraint, so the decay will take place
on a codimension-2 subspace of the restricted moduli space in which we
are working for this class of examples. The fact that in some
situations there are no solutions (for example if we do not satisfy
that $Q^{(1)'}_i$ and $Q^{(1)''}_i$ have opposite signs for
$i=3,4,5,6$) simply means that our restricted moduli space fails to
intersect the marginal stability locus in that case.

If the charges $Q^{(1)'}_i, Q^{(1)''}_i,P^{(1)'}_i, P^{(1)''}_i$
vanish for all $i\in 3,4,5,6$ then there are no constraints (beyond
the curve of marginal stability). It is easily seen that this is the
case where the final states are both $\half$-BPS.

\subsubsection*{(iii) General charges, ``diagonal'' moduli}

In this subsection we study rare decays allowing for completely
general charges $\vQ,\vP$, but we will restrict the moduli so that the
formulae are tractable. The situation turns out to be rather similar
to the case studied in the previous subsection.

Considerable simplification can be brought about in the formulae by
using some known results on T-duality orbits from Ref.\cite{Wall:1962}
(as reviewed in Appendix A of \cite{Banerjee:2007sr}).  For this
purpose we first change basis from the $L$ matrix used in
Ref.\cite{Banerjee:2007sr}:
\beq
L' = 
\begin{pmatrix}
\sigma_1 & 0 & \cdots & 0 & 0 & \cdots &0\\
0 & \sigma_1 & \cdots & 0 & 0 & \cdots &0 \\
0&& \cdots & 0 & 0 & \cdots &0 \\
0 & \cdots & 0 & \sigma_1 & 0 & \cdots &0 \\
0 && \cdots && - L_{E_8} && 0 \\
0 && \cdots && 0  && -L_{E_8} \\
\end{pmatrix}
\eeq
to the one we have defined in \eref{elldef}. Here $\sigma_1$ is a
Pauli matrix, which occurs 6 times in the above, and $L_{E_8}$ is the
Cartan matrix of $E_8$. 

In fact using T-duality we will be able to restrict to charge vectors
that have the last 16 components vanishing, therefore we can ignore
these components and work in a space of 12-component vectors. We then
use a $12\times 12$ matrix $X$ that satisfies
\beq
X L X^T = L'
\eeq
to map the equations in Ref.\cite{Banerjee:2007sr} to our basis.

Now the relevant result of T-duality orbits states that any pair of
primitive charge vectors $\vQ,\vP$ can be brought via T-duality to the
form:
\bea
\label{tdualcharges}
\vQ' &=& (Q'_1,0,\cdots,0),\qquad \vQ'' = (Q''_1,0,\cdots,0),\qquad
\vQ'''=0\nn\\ 
\vP' &=& (P'_1,P'_2,\cdots,0),\qquad \vP'' = (P''_1,P''_2,\cdots,0),\qquad 
\vP'''=0
\eea
This is close to our previous special case, but with $P'_1,P''_1$
turned on. It is no longer a special case but represents the general
case in a special basis.

As in the previous example, we restrict the moduli by requiring
$A^I_i=B_{ij}=0$ and $G_{ij}=0,~i\ne j$. Then one finds the projected
charges to be:
\beq
\vQ_R = 
\bmx
\frac{Q_1'}{R_1} + Q_1'' R_1\\
0\\ ...\\ 0
\emx, \qquad
\vP_R = 
\bmx
\frac{P_1'}{R_1} + P_1'' R_1
\\ \frac{P_2'}{R_2} + P_2'' R_2\\
0\\ ...\\ 0
\emx
\eeq

The quartic invariant is then found to be:
\beq
\Delta(Q_R,P_R) = \left(\frac{Q_1'}{R_1} + Q_1'' R_1\right)^2
\left(\frac{P_2'}{R_2} + P_2'' R_2\right)^2
\eeq
which is actually the {\it same} as in the previous, simpler case
where we chose a special subset of charges.  Computing
$m_1,r_1,s_1,n_1$ as in the previous subsection and inserting them
back, the constraint equation can now be written:
\bea
&&\left(\frac{Q_1'}{R_1} + Q_1'' R_1\right)
\left(\frac{P_2'}{R_2} + P_2'' R_2\right)\vQ^{(1)}_R =\nn\\
&&\Bigg[
\left(\frac{Q^{(1)'}_1}{R_1} + Q^{(1)''}_1 R_1\right)
\left(\frac{P_2'}{R_2} + P_2'' R_2\right)-
\left(\frac{Q^{(1)'}_2}{R_2} + Q^{(1)''}_2 R_2\right)
\left(\frac{P_1'}{R_1} + P_1'' R_1\right)\Bigg]\vQ_R +\nn\\
&&\qquad\qquad\qquad\qquad\qquad\qquad\qquad\qquad\qquad
\left(\frac{Q_1'}{R_1} + Q_1'' R_1\right)
\left(\frac{Q^{(1)'}_2}{R_2} + Q^{(1)''}_2 R_2\right)\vP_R\nn\\
&&\left(\frac{Q_1'}{R_1} + Q_1'' R_1\right)
\left(\frac{P_2'}{R_2} + P_2'' R_2\right) \vP^{(1)}_R =\nn\\
&&\Bigg[\left(\frac{P^{(1)'}_1}{R_1} + P^{(1)''}_1 R_1\right)
\left(\frac{P_2'}{R_2} + P_2'' R_2\right)-
\left(\frac{P^{(1)'}_2}{R_2} + P^{(1)''}_2 R_2\right)
\left(\frac{P_1'}{R_1} + P_1'' R_1\right)\Bigg]
\vQ_R +\nn\\
&&\qquad\qquad\qquad\qquad\qquad\qquad\qquad\qquad\qquad
\left(\frac{Q_1'}{R_1} + Q_1'' R_1\right)
\left(\frac{P^{(1)'}_2}{R_2} + P^{(1)''}_2 R_2\right)\vP_R\nn\\
\eea

These equations are slightly more complicated than the previous case
for which we had $\vQ\circ \vP=0$, but the extra complication is only
in the first two components, which are again trivially satisfied. For
the remaining components we find:
\bea
\frac{Q^{(1)'}_i}{R_i} + Q^{(1)''}_i R_i =0,\quad i=3,4,5,6\nn\\
\frac{P^{(1)'}_i}{R_i} + P^{(1)''}_i R_i =0,\quad i=3,4,5,6
\eea
These are exactly the {\it same} as the constraints we found in the
previous case. The analysis is therefore also the same: the
constraints cannot be satisfied for generic charges because our
restricted moduli space need not intersect the marginal stability
locus. When they can be satisfied there are at most four constraints,
though there will be less if some of the decay product charges vanish.

\subsection{General charges, ``triangular'' moduli}

In this subsection we restrict the moduli in the most minimal way
consistent with finding a simple form of the constraint equation. The
restriction will be a kind of ``triangularity'' condition:
\beq
(G + B+C)_{i1} = (G + B+C)_{i2} = 0, \quad i = 3,4,5,6
\eeq
with no separate constraints on $G,B,A$ other than the above.

As before, we use T-duality to put the initial charges into the form
of \eref{tdualcharges}. Thereafter, we are still free to make
T-duality transformations involving the last four components of $\vQ'$
and $\vQ''$ and all 16 components of $\vQ'''$. The T-duality group is
thus restricted to an $SO(4,20;\IZ)$. These transformations will
affect the charges of the decay products while leaving the initial
dyon unchanged. Using them we bring the electric charges of the first
decay product to the form:
\bea
\label{tdualprods}
\vQ^{(1)'} &=& (Q^{(1)'}_1,Q^{(1)'}_2,Q^{(1)'}_3,\cdots,0),\nn\\
 \vQ^{(1)''} &=& (Q^{(1)''}_1,Q^{(1)''}_2,Q^{(1)''}_3,
\cdots,0),\nn\\
\vQ^{(1)'''}&=&0
\eea
Finally we use an $SO(3,19;\IZ)$ subgroup of T-duality that preserves
all the charge vectors that we have so far fixed, to bring the
magnetic charges of the first decay product to the form:
\bea
\vP^{(1)'} &=&
(P^{(1)'}_1,P^{(1)'}_2,P^{(1)'}_3,P^{(1)'}_4,\cdots,0),\nn\\
\vP^{(1)''} 
&=& (P^{(1)''}_1,P^{(1)''}_2,P^{(1)''}_3,P^{(1)''}_4,\cdots,0),\nn\\
\vP^{(1)'''}&=&0
\eea
The charges of the second decay product are determined by charge
conservation.

Now we use the form of the projection matrix $\sqrt{L + M}$ and write
out \eref{lincomb} explicitly, after first multiplying through by
$E_{ij}$:
\bea
\label{lincombexp}
Q_i^{(1)'} + (G + B +C)_{ij} Q_j^{(1)''} &=& m_1
Q_i' + m_1 (G + B+C)_{ij}Q_j'' \nn\\ 
&&~~+\!\!\!\!\quad r_1 P_i' + r_1 (G + B+C)_{ij}P_k''\nn\\
P_i^{(1)'} + (G + B+C)_{ij}P_j^{(1)''} &=& s_1
Q_i' + s_1 (G + B+C)_{ij}Q_j'' \nn\\ 
&&~~+\!\!\!\! \quad n_1 P_i' + n_1 (G + B+C)_{ij} P_j''
\eea
This is a set of $6+6$ equations. Recall that $C_{ij}= A_i^I A_j^I$.

We immediately see that for our choice of T-duality frame for the
initial charges, as well as using the ``triangularity'' condition,
the RHS of the above equations vanishes for $i=3,4,5,6$.  Hence we
find the constraint equations still in a relatively simple form:
\bea
\label{constreqn}
Q_i^{(1)'} + 
(G+B+C)_{ij}Q_j^{(1)''} &=& 0,\quad
i=3,4,5,6\nn\\ 
P_i^{(1)'} + 
(G+B+C)_{ij}P_j^{(1)''} &=&
0,\quad i=3,4,5,6
\eea
These are then the 4+4 constraints on rare dyon decays, though still
with the triangularity restriction on moduli and in a specific
T-duality frame. They must be supplemented by the curve of marginal
stability, for which we need to know the numbers $m_1,r_1,s_1,n_1$.

The first two components of each line of equations \eref{lincombexp}
determine the values of $m_1,r_1,s_1,n_1$. From the first line of
those equations we find:
\bea
Q^{(1)'}_1 + (G + B+C)_{1i}\vect{Q}^{(1)''}_i &=& m_1 Q_1' + m_1 (G +
B+C)_{1i}Q_i'' + r_1 P_1' + r_1 (G + B+C)_{1i} P_i''\nn\\
Q^{(1)'}_2 + (G + B+C)_{2i} Q^{(1)''}_i &=& r_1 P_2' + r_1 (G +
B+C)_{2i} P_i''
\eea
Solving for $r_1$ from the second equation above, we get:
\beq
r_1 = \frac{Q^{(1)'}_2 + (G + B+C)_{2i}Q^{(1)''}_i}{P_2' + (G +
B+C)_{2i}P_i''}
\eeq
Inserting this in the first equation determines $m_1$:
\bea
m_1 &=& 
\Big(P_2' + (G + B+C)_{2i}P_i''\Big)^{-1}\Big(Q_1' + (G +
B+C)_{1i}Q_i''\Big)^{-1}\times\nn\\ 
&&
\Bigg[\Big(Q^{(1)'}_1 + (G + B+C)_{1i}Q^{(1)''}_i\Big)
\Big(P_2' + (G + B+C)_{2i}P_i''\Big) \nn\\
&& \qquad-\Big(Q^{(1)'}_2 + 
(G + B+C)_{2i}Q^{(1)''}_i\Big)\Big(P_1' 
+ (G + B+C)_{1i}P_i''\Big)\Bigg]
\eea

Similarly we solve for $s_1,n_1$ from the second line of
\eref{lincombexp} and find:
\bea
n_1 &=& \frac{P^{(1)'}_2 + (G + B+C)_{2i}P^{(1)''}_i}{P_2' + (G +
B+C)_{2i}P_i''}\nn\\
s_1 &=& \Big(P_2' + (G + B+C)_{2i}P_i''\Big)^{-1}
\Big(Q_1' + (G + B+C)_{1i}Q_i''\Big)^{-1}\times \nn\\
&&\Bigg[
\Big(P^{(1)'}_1 + (G + B+C)_{1i}P^{(1)''}_i\Big)\Big(P_2' + (G +
B+C)_{2i}P_i''\Big)\nn\\
&& \qquad - \Big(P^{(1)'}_2 + (G +
B+C)_{2i}P^{(1)''}_i\Big)\Big(P_1' + r (G +
B+C)_{1i}P_i\Big)\Bigg] 
\eea

Admittedly these are somewhat complicated expressions for the numbers
$m_1,r_1,s_1,n_1$ that one needs to plug in to determine the curve of
marginal stability on the torus moduli space. It is conceivable that a
more opportune choice of variables could simply them
further. Nevertheless, the constraints \eref{constreqn} on the
remaining moduli are rather simple.

\subsection{Explicit solution: the general case}

We now turn to the case where the initial and final charges are
completely general and the moduli are generic as well. Most of the
relevant analysis has already been done in previous subsections and it
only remains to write down the result. However, as we will see, the
equations rapidly become messy -- despite the use of T-duality
transformations - once we use completely general moduli.

Let us again start by writing out \eref{lincomb} explicitly, but now
without any condition on the moduli. After multiplying through by
$E_{ij}$, we find the equations:
\bea
\label{lincombexpgen}
Q_i^{(1)'} + (G + B +C)_{ij} Q_j^{(1)''} &=& m_1
Q_i' + m_1 (G + B+C)_{ij}Q_j'' \nn\\ 
&&~~+\!\!\!\!\quad r_1 P_i' + r (G + B+C)_{ij}P_k''\nn\\
P_i^{(1)'} + (G + B+C)_{ij}P_j^{(1)''} &=& s_1
Q_i' + s_1 (G + B+C)_{ij}Q_j'' \nn\\ 
&&~~+\!\!\!\! \quad n_1 P_i' + n_1 (G + B+C)_{ij} P_j''
\eea
which are actually the same as \eref{lincombexp} that we had
before. The difference is that the RHS no longer vanishes for any of
the components (earlier that was guaranteed by the triangularity
condition that we had assumed on the moduli). Notice that even in the
most general case, we have gained something by fixing the initial and
final state charges using T-duality. The last 16 components of these
charges have all been set to 0, and the result is that most of the
terms involving the gauge field moduli $A_i^I$ have disappeared. The
only appearance of these moduli is through $C_{ij}=A_i^IA_j^I$ which
in turn only appears in the combination $G+B+C$.

This time our strategy will be to choose any 4 equations from the
above set of 12 to determine the variables $m_1,n_1,r_1,s_1$. Then in
the remaining 8 equations we insert these values for the variables and
obtain the desired constraint equations. Picking the first 2
components for each charge vector, we find:
\bea
Q^{(1)'}_1 + (G + B+C)_{1i}Q^{(1)''}_i &=& m_1 Q_1' + m_1 (G +
B+C)_{1i} Q_i'' + r_1 P_1' + r (G + B+C)_{1i} P_i''\nn\\
Q^{(1)'}_2 + (G + B+C)_{2i} Q^{(1)''}_i &=& r_1 P_2' + r_1 (G +
B+C)_{2i} P_i''
\eea
Solving for $r_1$ from the second equation above, we get:
\beq
r_1 = \frac{Q^{(1)'}_2 + (G + B+C)_{2i}Q^{(1)''}_i}{P_2' + (G +
B+C)_{2i}P_i''}
\eeq
and inserting this in the first equation, we find $m_1$:
\bea
m_1 &=& \Big(P_2' + (G + B+C)_{2i}P_i''\Big)^{-1}
\Big(Q_1' + (G + B+C)_{1i} Q_i''\Big)^{-1}\times\nn\\
&&\Bigg[\Big(Q^{(1)'}_1 + (G + B+C)_{1i} Q^{(1)''}_i\Big)
\Big(P_2' + (G + B+C)_{2i}P_i''\Big) \nn\\
&&\qquad - \Big(Q^{(1)'}_2 + (G +
B+C)_{2i}\vect{Q}^{(1)''}_i\Big)
\Big(P_1' + r (G + B+C)_{1i} P_i\Big)\Bigg]
\eea
Similarly we solve for $s_1,n_1$ from the second equation and find:
\beq
n_1 = \frac{P^{(1)'}_2 + (G + B+C)_{2i}P^{(1)''}_i}{P_2' + (G +
B+C)_{2i}P_i''}
\eeq
and
\bea
s_1 &=& \Big(P_2' +(G + B+C)_{2i}P_i''\Big)^{-1}\Big(Q_1' + (G +
B+C)_{1i}Q_i''\Big)^{-1}\times\nn\\
&& \Bigg[\Big(P^{(1)'}_1 + (G + B+C)_{1i}P^{(1)''}_i\Big)
\Big(P_2' + (G + B+C)_{2i}P_i''\Big)\nn\\
&&\quad - \Big(P^{(1)'}_2 + (G +
B+C)_{2i} P^{(1)''}_i\Big)\Big(P_1' + (G + B+C)_{1i}P_i\Big)
\Bigg] 
\eea
We feed in these values of $m_1,n_1,r_1,s_1 $ into the remaining 8
equations to find the most general constraint equations on the moduli:
\bea
Q^{(1)'}_i + (G + B+C)_{ij}Q^{(1)''}_j 
&=& m_1 \Big(Q_i' + (G + B+C)_{ij}Q_j''\Big) +  
r_1 \Big(P_i' + (G + B+C)_{ij}P_j''\Big)\nn\\
P^{(1)'}_i + (G + B+C)_{ij}P^{(1)''}_j 
&=& s_1 \Big(Q_i' + (G + B+C)_{ij}Q_j''\Big) 
+ n_1 \Big(P_i' + (G + B+C)_{ij}P_j''\Big)\nn\\
\eea
here $i = 3,4,5,6$, and $m_1,n_1,r_1,s_1 $ are given in the above
equations. We see that the values of $m_1,n_1,r_1,s_1 $ come out the
same as in the previous special case, however the constraints are much
more complicated and -- unlike in all the previous special cases --
depend explicitly on these numbers. Nevertheless, the above equations
embody the most general kinematic constraints on moduli space to allow
a two-body decay of a dyon of charges $\vQ,\vP$ into $\frac14$-BPS
final state with charges $\vQ^{(1)}$ and $\vP^{(1)}$ (the charges of
the second state being, as always, determined by charge
conservation). It is quite conceivable that a more detailed study of
possible T-duality bases will allow us to further simplify the most
general case, and we leave such an investigation for the future.

\section{Multi-particle decays}

So far in this work, as well as in previous
work\cite{Mukherjee:2007nc}, we have written down conditions for decay
of a dyon into two $\frac14$-BPS final states. One could certainly
imagine extending these considerations to three or more final
states. Indeed, it turns out rather simple to do so and we will here
discuss an iterative way to obtain the relevant formulae.

Consider the decay of a dyon of charges $(\vQ,\vP)$ into $n$ decay
products of charges $(\vQ^{(1)},\vP^{(1)}),
(\vQ^{(2)},\vP^{(2)}),\cdots (\vQ^{(n)},\vP^{(n)})$. The condition for
marginality of such a decay is the condition for the original dyon to
go into two decay products of charges $(\vQ^{(1)},\vP^{(1)})$ and
$\sum_{i=2}^n(\vQ^{(i)},\vP^{(i)})$, along with the condition for the
second decay product to further decay into say $(\vQ^{(2)},\vP^{(2)})$ and
$\sum_{i=3}^n(\vQ^{(i)},\vP^{(i)})$. The latter
condition must in turn be iterated. Each of these is a two-body decay
(with both final states being $\frac14$-BPS) so we already know the
condition for each one to take place. The intersection of all these
loci will give the marginal stability locus for the multiparticle decay.

There is a simpler way to iterate the condition. Instead of looking at
the curve where the second decay product decays into further
subconstituents, as above, we can simply consider the collection of
all marginal stability loci for the decays:
\beq
\ncr{\vQ_R}{\vP_R}\to
\ncr{\vQ^{(i)}_R}{\vP^{(i)}_R} +
\ncr{\vQ_R-\vQ^{(i)}_R}{\vP_R-\vP^{(i)}_R},\quad i=1,2,\cdots,n
\eeq
For each of these, the curve is precisely \eref{cms} with the
subscript ``$1$'' replaced by ``$i$''. We write it as:
\beq
\label{cmsi}
{\cal C}(m_i,r_i,s_i,n_i)\equiv
\left(\tau_1-\frac{m_i- n_i}{2s_i}\right)^2
+ \left(\tau_2 + \frac{E_i}{2s_i}\right)^2
- \frac{1}{4s_i^2}\Big((m_i-n_i)^2 + 4r_is_i + E_i^2\Big)=0
\eeq
where
\beq
E_i\equiv \frac{1}{\sqrt\Delta}\left(\vQ^{(i)}\circ \vP - \vP^{(i)}\circ
\vQ\right)
\eeq
In addition to this curve we have the constraints on the remaining
moduli as in Sec.3 above. Those too can be expressed in terms of the
single decay product labelled ``$i$''. Now to find the condition for a
multi-dyon decay, we simply take the intersection of all these loci of
marginal stability. As the number of final states increases, we will
generically find loci of marginal stability of increasing codimension.

\section{Multi-centred black holes}

It was argued in Refs.\cite{Sen:2007pg,Cheng:2007ch} that the curves
of marginal stability for decays of the form:
\beq
\sfrac14\hbox{-BPS}\to\sfrac12\hbox{-BPS}+\sfrac12\hbox{-BPS}
\eeq
are also the curves of disintegration for two-centred $\frac14$-BPS
black holes whose centres are individually $\frac12$-BPS. The method
used in these works, which we will summarise and extend below, was to
use a constraint equation due to Denef \cite{Denef:2000nb} to
express the separation between the centres of such a black hole in
terms of charges and moduli. Requiring that the separation be infinite
places a condition on charges and moduli which turns out to be
precisely the curve of marginal stability, \eref{cms}, specialised to
this decay.

Now Denef's constraint equation is not confined to
two-centred black holes alone, but applies to any number of
centres. It has a different limitation: it is defined in the context
of ${\cal N}=2$, rather than ${\cal N}=4$ compactifications, and
relies on special geometry. Nevertheless, for the cases to which it
applies, we can certainly use it in the ${\cal N}=4$ context. We will
do so and will find the perhaps surprising result that the curves of
marginal stability for generic decays to $n$ final states, which we
discussed in Sec.4 above, are precisely reproduced by the constraint
equations for multi-centred black holes. This suggests a more generic
relationship between multi-particle decays and multi-centred black
holes than has been previously considered.  

The constraint equation on multi-centred dyons, (see for example
Ref.\cite{Sen:2007pg}\footnote{A sign in equation (3.2) of
Ref.\cite{Sen:2007pg} should be corrected so that it reads
$\frac{X^1}{X^0} = -\tau$. This leads to some sign changes in other
equations there.}) reads as follows. Let $p^{(i)I},q^{(i)}_I$ be the
charges of the $i$-th centre where $i=1,2,\cdots,N$. These charges are
expressed in the special-geometry basis\footnote{As we will see, this
differs by an interchange of some components from the standard basis
used in ${\cal N}=4$ compactifications.}. Let the 3-vector ${\vec
r}_i$ be the location of the $i$-th centre. And let the moduli be
encoded in the standard holomorphic special-geometry variables
$X^I,F_I$. Then the constraint equations are:
\beq
\label{constreq}
p^{(i)I}\Im(F_{I\infty}) - q^{(i)}_I\Im(X^I_\infty) +
\half \sum_{j\ne i} 
\frac{p^{(i)I}q^{(j)}_I -q^{(i)}_I p^{(j)I}}{|\vr_i-\vr_j|}=0
\eeq
Here the subscript $\infty$ indicates that the corresponding moduli
are measured at spatial infinity (for brevity of notation we will drop
it when there is no risk of ambiguity). Note that the numerators inside
the summation correspond to the Saha angular momentum between each
pair of centres.

These are $N$ equations for $({N\atop 2})$ pairwise distances between
the centres. We analyse them following the procedure in
Ref.\cite{Sen:2007pg} for the two-centred case.  First of all, one of
the equations is redundant. Adding all the equations, we find:
\beq
\label{totaleq}
p^I \Im(F_{I\infty}) - q_I \Im(X^I_\infty) =0
\eeq
where $(p^I,q_I)$ are the charges of the entire black hole. This
provides one real constraint on the extra modulus $X^0_\infty$. As the
above equation is invariant under $X^I\to \lambda X^I$ for real $X^I$,
as well as under $X^I\to -X^I$, we see that the magnitude of $X^0$ is
undetermined by this condition, while the phase is determined (in
terms of the $X^I,I=1,2,3$) upto a two-fold ambiguity. Another real
constraint is now imposed in the form of a ``gauge condition'':
\beq
\label{gaugecon}
X^I \bF_I - \bX^I F_I =-i
\eeq
This determines the magnitude of $X^0$ but leaves intact the two-fold
ambiguity in the phase. The remaining $N-1$ equations then provide
constraints on the $({N\atop 2})$ separations. 

For the case $N=2$ we therefore have a single equation, which
completely determines the separation between the two centres. This
works as follows. The relevant part of the theory is described by the
holomorphic prepotential:
\beq
F = - \frac{X^1 X^2 X^3}{X^0} 
\eeq
where the $X^I$ are complex scalar fields related to a subset of the
$K3\times T^2$ moduli, namely $\tau=\tau_1+i\tau_2$ describing the
2-torus complex structure, and
\beq
M = {\rm diag}({\hat R}^{-2},R^{-2},{\hat R}^2, R^2)
\eeq
describing a 2-parameter subset of the remaining moduli (including the
$K3$ moduli). The precise relationship is:
\beq
\frac{X^1}{X^0} = -\tau,\quad \frac{X^2}{X^0} = iR\hat{R},\quad
\frac{X^3}{X^0} = i\frac{\hat{R}}{R}
\eeq
The gauge condition \eref{gaugecon} then tells us that:
\beq
\label{gaugecon2}
|X^0_\infty|^2 = \frac{1}{8 \hat{R}^2 \,\tau_2}
\eeq

As in the previous sections, we will consider a dyon with charges
$(\vQ,\vP)$, but now each taken to be 4-component (the first two
components should be thought of as two of the six $\vQ'$ and the
second two components constitute two of the six $\vQ''$. The charges
correspond to unit torsion, namely:
\beq
{\rm g.c.d.}(Q_iP_j- P_iQ_j)=1
\eeq
We begin by determining the modulus $X^0$ in terms of the T-duality
invariants $P\circ P,Q\circ Q,P\circ Q$, where as before the inner
products are defined in terms of the moduli at infinity, e.g. $P\circ
P = P^T(L+M)P$. 

As promised, we will use the transcription between the natural
electric-magnetic basis $\vP,\vQ$ for the type IIB superstring and the
natural basis $p^I,q_I$ for special geometry (see for example Ref.
\cite{Sen:2007pg}):
\beq
\label{chargerel}
q_I=(Q_1,P_1,Q_4,Q_2),\quad
p^I=(P_3,-Q_3,P_2,P_4)
\eeq
In addition we have:
\beq
\begin{split}
\Im(F_0)&={\hat R}^2 \Im(X^0\tau),\quad
\Im(F_1)={\hat R}^2 \Im(X^0),\\
\Im(F_2)&=\frac{\hat R}{R} \Re(X^0\tau),\quad
\Im(F_3)=R{\hat R} \Re(X^0\tau)\\
\end{split}
\eeq
while
\beq
\begin{split}
\Im(X^0)&=\Im(X^0),\quad
\Im(X^1)=-\Im(X^0\tau),\\
\Im(X^2)&= R{\hat R}\Re (X^0),\quad
\Im(X^3)= \frac{\hat R}{R}\Re (X^0)\\
\end{split}
\eeq
Inserting these into Eqs.(\ref{totaleq}),(\ref{gaugecon2}), one finds:
\beq
X^0 = \frac{1}{(2\sqrt{2} \hat{R} \tau_2)}  
\frac{\sqrt{\Delta} \btau 
+i\left(Q\circ P \btau - Q\circ Q\right)}{\sqrt{Q\circ Q}\, M_{BPS}}
\eeq
where $M_{BPS}$ is the BPS mass given by
\eref{genbpsmass}.

Now let us assume our dyon has $n$ centres of charges
$(\vQ^{(i)},\vP^{(i)})$:
\beq
\label{chargeassign}
\ncr{\vQ^{(i)}}{\vP^{(i)}}= \pmat{m_i}{r_i}{s_i}{n_i}
\ncr{\vQ}{\vP},\quad i=1,2,\cdots,n
\eeq
with $m_i,r_i,s_i,n_1$ integers satisfying:
\beq
\sum_{i=1}^n m_i= \sum_{i=1}^n n_i=1,\quad \sum_{i=1}^n r_i
= \sum_{i=1}^n s_i =0
\eeq

From \eref{chargerel} we find that the charges of the decay products
in the $q_I,p^J$ basis are given by:
\beq
\begin{split}
q^{(i)}_I&= (m_i Q_1 + r_iP_1,s_i Q_1+n_iP_1, m_iQ_4 + r_i P_4,
m_i Q_2 + r_i P_2)\\
p^{(i)I} &= (s_i Q_3 + n_i P_3, -(m_i Q_3 + r_i P_3), s_i Q_2 + n_i
P_2, s_i Q_4 + n_i P_4)\\
\end{split}
\eeq
Now the first term in \eref{constreq} can be written:
\beq
\label{firstterm}
p^{(i)I}\Im(F_I) - q^{(i)}_I\Im(X^I) = {\hat R} 
\Re\begin{pmatrix}
-X^0 & X^0\tau
\end{pmatrix}
\pmat{m_i}{r_i}{s_i}{n_i}
\ncr{\frac{Q_2}{R} + R Q_4 -i(\frac{Q_1}{\hat R}+ {\hat R}Q_3)}
{\frac{P_2}{R} + R P_4 -i(\frac{P_1}{\hat R}+ {\hat R}P_3)}
\eeq

The invariants $P\circ P, Q\circ Q, Q\circ P$ are given by:
\beq
\begin{split}
Q\circ Q &= \left(\frac{Q_1}{{\hat R}} + {\hat R}Q_3\right)^2+
\left(\frac{Q_2}{R} + R Q_4\right)^2\\
P\circ P &= \left(\frac{P_1}{{\hat R}} + {\hat R}P_3\right)^2+
\left(\frac{P_2}{R} + R P_4\right)^2\\
Q\circ P &= \left(\frac{Q_1}{{\hat R}} + {\hat R}Q_3\right)
\left(\frac{P_1}{{\hat R}} + {\hat R}P_3\right) +
\left(\frac{Q_2}{R} + R Q_4\right)\left(\frac{P_2}{R} + R P_4\right)
\\
\end{split}
\eeq
The column vector in \eref{firstterm} depends on four combinations of
$Q_i,P_i$ and therefore cannot in general be expressed in terms of
T-duality invariants. Therefore we restrict to the special case,
discussed in particular in Ref.\cite{Sen:2007pg}, for which
$Q_1=Q_3=0$. In this case only three independent combinations appear
in the column vector and it is easy to show that:
\beq
\begin{split}
p^{(i)I}\Im(F_I) - q^{(i)}_I\Im(X^I) &= {\hat R} 
\Re X^0\begin{pmatrix}
-1 &\tau
\end{pmatrix}
\pmat{m_i}{r_i}{s_i}{n_i}
\ncr{\sqrt{Q\circ Q}}
{\frac{Q\circ P + i\sqrt\Delta}{\sqrt{Q\circ Q}}}\\[2mm]
&= \frac{s_1\sqrt{\Delta}}{2\sqrt2\,\tau_2\, M_{BPS}}~
{\cal C}(m_i,r_i,s_i,n_i)
\\
\end{split}
\eeq
where ${\cal C}(m_i,r_i,s_i,n_i)$ is the curve of marginal stability
for multiparticle decays, defined in \eref{cmsi}.

The numerator of the second term in \eref{constreq}, denoted:
\beq
{\cal J}_{ij}~\equiv~ p^{(i)I}q^{(j)}_I -p^{(j)I} q^{(i)}_I
\eeq
is the angular momentum between each pair of decay products
evaluated in the moduli-independent norm. We will denote the
pairwise separation between the centres by:
\beq
L_{ij}=|\vr_i-\vr_j|
\eeq
Note that $\cJ_{ij}=-\cJ_{ji}$ and $L_{ij}=L_{ji}$.  

Inserting the above results into \eref{constreq}, one finds that it
can be expressed as follows:
\beq
\label{constrsimp}
\tcC_i +
\sum_{j\ne i} \frac{{\cal J}_{ij}}{L_{ij}}=0
\eeq
where
\beq
\tcC_i \equiv
\frac{s_i\,\sqrt{\Delta}}{\sqrt2\,\tau_2\, M_{BPS}}~
{\cal C}(m_i,r_i,s_i,n_i)
\eeq
Clearly the first term in \eref{constrsimp} depends only on the
charges of a single centre (as well as the initial charges) while the
second term depends on the charges of a pair of centres. Note that we
have $\sum_i \tcC_i =0$. Thus we have shown that the curves of
marginal stability for multi-centred decays appear also from
considerations of multi-centred black holes and the constraints on the
locations of their centres. 

In the special case considered previously
\cite{Sen:2007pg,Cheng:2007ch} where the dyon has two $\half$-BPS
centres, the corresponding curve of marginal stability is of
codimension 1. In this case it is known that the degeneracy of states
jumps as we cross the curve. From the
supergravity point of view, it was suggested in the ${\cal N}=2$
context in Ref.\cite{Denef:2000nb} and shown more explicitly in the
present ${\cal N}=4$ context in Refs.\cite{Sen:2007pg,Cheng:2007ch},
that this decay occurs as a result of the two centres flying apart to
infinity at a curve of marginal stability. This is seen by
specialising
\eref{constrsimp} to this case. As long as $\cJ_{12}\ne
0$, the separation $L_{12}\to\infty$ when $\tcC_i\to 0$. Moreover for
a fixed sign of $\cJ_{12}$, the separation $L_{12}$ can be positive
only on one side of the curve of marginal stability. On the other side
it is negative, which indicates that the corresponding two-centred
black hole does not exist.

Now let us return to the more general case where there are two centres
but both are $\frac14$-BPS. As we have seen, in this case the locus of
marginal stability is not a wall in moduli space, but rather a curve
of codimension $\ge 2$. Therefore the degeneracy formula cannot jump
as one crosses the curve. Hence one need not have expected any
relationship between marginal decays and multi-centred
dyons. Nevertheless, we see that \eref{constrsimp} continues to hold
in the more general case (with the limitation that the charges are
those that can be embedded in an ${\cal N}=2$ compactification). 

We interpret this as evidence that the relationship between dyon decay
and the disintegration of multi-centred black holes holds more
generally than required by the degeneracy formula.  Therefore we
conjecture that even with the most general charges, $n$-centred black
holes exist in ${\cal N}=4$ string compactifications with generic
$\frac14$-BPS centres for which \eref{constrsimp} holds true. It would
be worth trying to prove that this is the case, or else to show that
such solutions do not exist beyond the cases that can be embedded in
the charge space and moduli space of ${\cal N}=2$. An intermediate
possibility also exists: that in ${\cal N}=4$ compactifications such
multi-centred black holes do exist with arbitrary charges, but only on
a subspace of the moduli space.

Examining \eref{constrsimp} one sees that if the marginal stability
condition $\tcC_i=0$ is satisfied for a particular $i$, then we must
have: 
\beq
\sum_{j\ne i}\frac{\cJ_{ij}}{L_{ij}}=0
\eeq
One possible solution is to have $L_{ij}\to\infty$ for all $j\ne
i$. This means the $i$th centre has been taken infinitely far away
from all the others, in agreement with the picture of marginal decay
that we developed in Section 4 above. Since the pairwise Saha angular
momenta $\cJ_{ij}\equiv P^{(i)}\cdot Q^{(j)} - P^{(j)}\cdot Q^{(i)}$
cannot all be positive in every equation (since $\cJ_{ij}=-\cJ_{ji}$)
there could be other configurations where the $\tcC_i=0$, except in
the case of two centres. It is not clear to us how these other
solutions should be interpreted.

Note that in the above equation the angular momentum is measured with
respect to moduli-independent inner product $P\cdot Q \equiv P^T L Q$
unlike the angular momentum appearing in the curve of marginal
stability \eref{cmsi} which is computed using the moduli-dependent
inner product $P\circ Q\equiv P^T(L+M)Q$. One may think of the latter
evolving to the former as we follow the attractor flow from infinity
to the horizon of the black hole. However it would be nice to have a
better physical understanding of the role of dyonic angular momenta in
these discussions\footnote{As is well-known, the dyonic angular
momentum plays a physical role in the wall-crossing formulae
\cite{Denef:2000nb,Denef:2007vg,Sen:2007pg,Cheng:2007ch,
Diaconescu:2007bf,Cheng:2008fc} that describe how the degeneracy
jumps, but in the present discussions there are no walls or jumps.}.

\section{Conclusions}

In this work we have obtained the loci of marginal stability for
decays of $\frac14$-BPS dyons into any number of BPS constituents in
${\cal N}=4$ string compactifications. These loci appear as equations
constraining the $132+2$ moduli, more precisely as a curve of marginal
stability in the upper-half-plane that represents a torus moduli space
(in the basis of type IIB on $K3\times T^2$, this is the geometric
torus) as well as some more complicated equations on the remaining
moduli. While in this paper we worked with unit-torsion initial dyons,
it should be quite straightforward to extend our results to general
torsion. We showed how to extend our analysis to multi-particle
decays, and found a relation between the loci of marginal stability
obtained in this way and the supergravity constraints on pairwise
separations of the centres of multi-centred black holes.

The physical role of ``rare'' marginal dyon decays, namely all those
other than of a $\frac14$-BPS dyon into two $\half$-BPS dyons, has yet
to be explored. Because such decays take place on loci of codimension
$\ge 2$ in moduli space, they do not form ``domain walls'' across
which the degeneracy can jump. Therefore they do not affect the basic
entropy or dyon counting formulae. However it is certainly possible
that they have other interesting physical effects which may emerge on
further investigation.

\section*{Acknowledgements}

We would like to thank Anindya Mukherjee, Suresh Nampuri and Ashoke
Sen for very helpful discussions. The work of S.M. was supported in
part by a J.C. Bose Fellowship of the Government of India. We are
grateful to the people of India for generously supporting research in
string theory.

\bibliographystyle{JHEP}

\bibliography{constraints}

\end{document}